\documentclass{iau}
\usepackage{graphicx} 

\title[IAUS291.~~Recycling Pulsars: spins, masses and ages] 
{Recycling Pulsars: spins, masses and ages} 

\author[T.M.~Tauris, M.~Kramer \& N.~Langer]  
{T.M.~Tauris$^{1,2,*}$, M. Kramer$^2$
 \and N.~Langer$^1$}

\affiliation{$^1\,$Argelander-Institut f\"ur Astronomie, Universit\"at Bonn, Germany\\[\affilskip] 
             $^2\,$Max-Planck Institut f\"ur Radioastronomie, Bonn, Germany \\[\affilskip]
             $^{*}\,$email: {\tt tauris@astro.uni-bonn.de} \\
}

\pubyear{2012}
\volume{291}  
\jname{\mbox{Neutron Stars and Pulsars: Challenges and Opportunities after 80 years}}
\editors{J.~van Leeuwen, ed.} 
\begin{document}

\maketitle

\begin{abstract}
Although the first millisecond pulsars (MSPs) were discovered 30 years ago we
still do not understand all details of their formation process. Here,
we present new results from \cite[Tauris,~Langer~\& Kramer~(2012)]{tlk12} on the recycling scenario leading to radio
MSPs with helium or carbon-oxygen white dwarf companions
via evolution of low- and intermediate mass X-ray binaries (LMXBs, IMXBs). 
We discuss the location of the spin-up line in the $P\dot{P}$--diagram
and estimate the amount of accreted mass needed to obtain a given spin period and
compare with observations. Finally, we constrain the true ages of observed
recycled pulsars via calculated isochrones in the $P\dot{P}$--diagram.

\end{abstract}

\firstsection 
\section{Introduction}
Binary MSPs represent the advanced phase of stellar evolution in close, interacting binaries. Their observed orbital
and stellar properties are fossil records of their evolutionary history. Thus one
can use binary pulsar systems as key probes of stellar astrophysics.
Although the standard recycling scenario (\cite[Alpar~et~al.~1982]{acrs82}; \cite[Bhattacharya~\&~van~den~Heuvel~1991]{bv91}) is commonly accepted,
many aspects of the mass-transfer process and the accretion physics are
still not understood in detail. Examples of such ambiguities include the accretion disk structure,
the disk-magnetosphere transition zone, the accretion efficiency,
the decay of the surface B-field of the neutron star
and the outcome of common envelope evolution. For further details on these aspects, details in general and discussions
of our results, we refer to
our journal paper, \cite[Tauris~et~al.~(2012)]..

\section{The spin-up line}
 \begin{figure}[b]
 \begin{center}
  \includegraphics[width=3.4in, angle=-90]{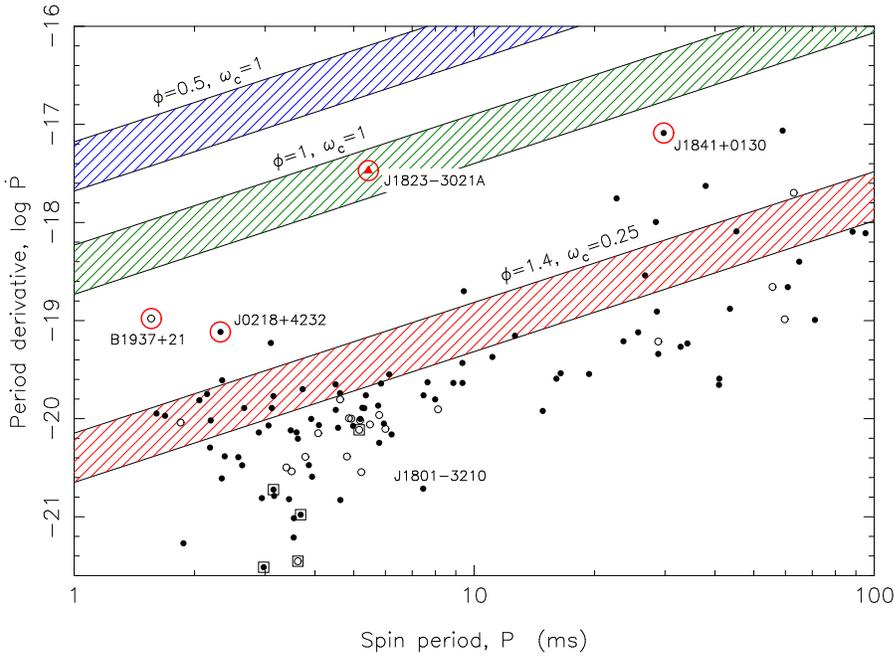} 
  \caption{Calculations of three spin-up lines, 
     shown as coloured bands, depending on the parameters ($\phi$,~$\omega_{\rm c}$).
     The upper boundary of each band (or ``line'')
     is calculated for a neutron star mass $M=2.0\,M_{\odot}$ and $\alpha = 90^{\circ}$.
     The lower boundary is calculated for
     $M=1.0\,M_{\odot}$ and $\alpha = 0^{\circ}$.
     The green (central) hatched band corresponds to $\phi=1$ and $\omega_{\rm c}=1$.
     The blue and red hatched bands are upper and lower limits set by reasonable choices
     of the two parameters ($\phi$,~$\omega_{\rm c}$).
     In all three cases the spin-up line is calculated assuming accretion at the Eddington limit, $\dot{M}=\dot{M}_{\rm Edd}$.
     The observed distribution of binary and isolated radio pulsars in the Galactic disk
     are plotted as filled and open circles, respectively.
     Also plotted is the pulsar J1823$-$3021A, located in the globular cluster NGC~6624. 
     (Fig. adapted from Tauris et al.\ 2012.)
    } 
    \label{fig1}
 \end{center}
 \end{figure}
Some of the above mentioned simplifications become a problem when trying to probe the formation and the evolution of observed
recycled radio pulsars located near the classical spin-up line for Eddington accretion in the $P\dot{P}$--diagram
(e.g. as illustrated with the MSP J1823$-$3021A, \cite[Freire~et~al.~2011]{faa+11}).
The location of the spin-up line can be found by
considering the equilibrium configuration when the angular velocity of the neutron star is equal to the
Keplerian angular velocity of matter at the magnetospheric boundary ($r_{\rm mag}$) where the accreted matter enters the magnetosphere,
i.e. $\Omega _\star = \Omega _{\rm eq} = \omega _c\,\Omega _{\rm K}(r_{\rm mag})$ or: 
$P_{\rm eq} = 2\pi (r_{\rm mag}^3/GM)^{1/2}\,\omega _c^{-1}$,
where $0.25 < \omega _c \le 1$ is the so-called critical fastness parameter. Introducing the magnetospheric
coupling radius, $\phi \equiv r_{\rm mag}/r_{\rm Alfven}$ and the magnetic inclination angle, $\alpha$ we can rewrite this expression:
\begin{equation}
     \dot{P} = \frac{2^{1/6}G^{5/3}}{\pi ^{1/3} c^3}\frac{\dot{M}M^{5/3}P_{\rm eq}^{4/3}}{I}
                \;\,(1+\sin^{2}\alpha)\;\,\phi^{-7/2}\,\omega_c^{7/3} 
\label{eq:spinupline} 
\end{equation}
which can be plotted directly in the $P\dot{P}$--diagram. 
(Here $M$ is the mass of the pulsar, $\dot{M}$ is its accretion rate and $I$ is its moment of inertia.) In case $\sin \alpha = \phi = \omega _c =1$ we find:
\begin{equation} 
     \dot{P} = 3.7\times 10^{-19}\,\;(M/M_{\odot})^{2/3}\,P_{\rm ms}^{4/3} \left(\frac{\dot{M}}{\dot{M}_{\rm Edd}}\right) 
  \label{eq:spinuplinefit2}
\end{equation}
where $P_{\rm ms}$ is the equilibrium spin period in units of milliseconds. We have 
included the plasma term in the spin-down torque using the combined model of \cite[Spitkovsky~(2006)]{spi06} to
compensate for the incompleteness of the vacuum magnetic dipole model.

In Fig.~\ref{fig1} we have plotted equation~(\ref{eq:spinupline}) for different values of $\alpha$, $\phi$ and $\omega_{\rm c}$
to illustrate the uncertainties in the applied accretion physics to locate the spin-up line.
In all cases we assumed a fixed accretion rate of $\dot{M}=\dot{M}_{\rm Edd}$. The location of the spin-up line
is simply shifted one order of magnitude in $\dot{P}$ down (up) for every order of magnitude $\dot{M}$
is decreased (increased).
It is important to realize that there is no universal spin-up line in the $P\dot{P}$--diagram. Not only are $M$ and $\alpha$ individual to each pulsar
(giving rise to the width of each band),
also $\phi$ and $\omega _c$ could be related to $B$, $\alpha$ and $\dot{M}$.\\
If we assume that accretion onto the neutron star is indeed Eddington limited,
then the three bands in Fig.~\ref{fig1} represent upper limits for the spin-up line for the
given sets of ($\phi$,~$\omega _{\rm c}$).
Thus we can in principle use this plot to constrain ($\phi$,~$\omega _{\rm c}$) and hence the physics of disk--magnetosphere
interactions from future detections of MSPs.

When modeling the birth spins of recycled radio pulsars it is important to include the braking torque acting
during the Roche-lobe decoupling phase (RLDP) when the donor star terminates its mass transfer.
It has been shown that accreting X-ray MSPs may lose up to 50~\% of their rotational energy
during the RLDP of LMXBs (\cite[Tauris~2012]{tau12}).

\section{Relation between accreted mass and final spin period}
\label{sec:mass-spin}
Recycled pulsars obtain their fast spins via angular momentum exchange from the differential rotation between the accretion disk
and the neutron star.
The amount of spin angular momentum added to an accreting pulsar is given by:
\begin{equation}
  \Delta J_\star = \int n(\omega,t)\,\dot{M}(t)\,\sqrt{GM(t)r_{\rm mag}(t)}\,\xi (t)\;dt
  \label{eq:Jacc}
\end{equation}
where $n(\omega,t)$ is a dimensionless torque.
Assuming $n(\omega,t)=1$, and $M(t)$, $r_{\rm mag}(t)$ and $\xi(t)$ to be roughly constant during the major part of the spin-up phase
we can obtain a simple and 
convenient expression to relate the (minimum) amount of accreted mass 
and final equilibrium spin period (see also \cite[Alpar~et~al.~1982]{arcs82}):
\begin{equation} 
     \Delta M_{\rm eq} = 0.22\,M_{\odot}\; \frac{(M/M_{\odot})^{1/3}}{P_{\rm ms}^{4/3}}
  \label{eq:deltaMfinalfit}
\end{equation}
assuming a numerical factor $f(\alpha,\xi,\phi,\omega_c)=1$ (from disk-magnetosphere interactions).\\
Considering a pulsar with a final mass of $1.4\,M_{\odot}$ and a recycled spin period
of either 2~ms, 5~ms, 10~ms or 50~ms requires accretion of
$0.10\,M_{\odot}$, $0.03\,M_{\odot}$, $0.01\,M_{\odot}$ or $10^{-3}\,M_{\odot}$, respectively.
Therefore, it is no surprise that observed recycled pulsars with massive companions ({CO}/{ONeMg}~WD or NS)
in general are much more slow rotators -- compared to MSPs with {He}~WD companions --
since the progenitor of their massive companions evolved on a relatively short timescale in IMXBs or HMXBs,
only allowing for very little mass to be accreted by the pulsar.

\section{True age isochrones of recycled pulsars}
 \begin{figure}[t]
 \begin{center}
  \includegraphics[width=3.7in, angle=-90]{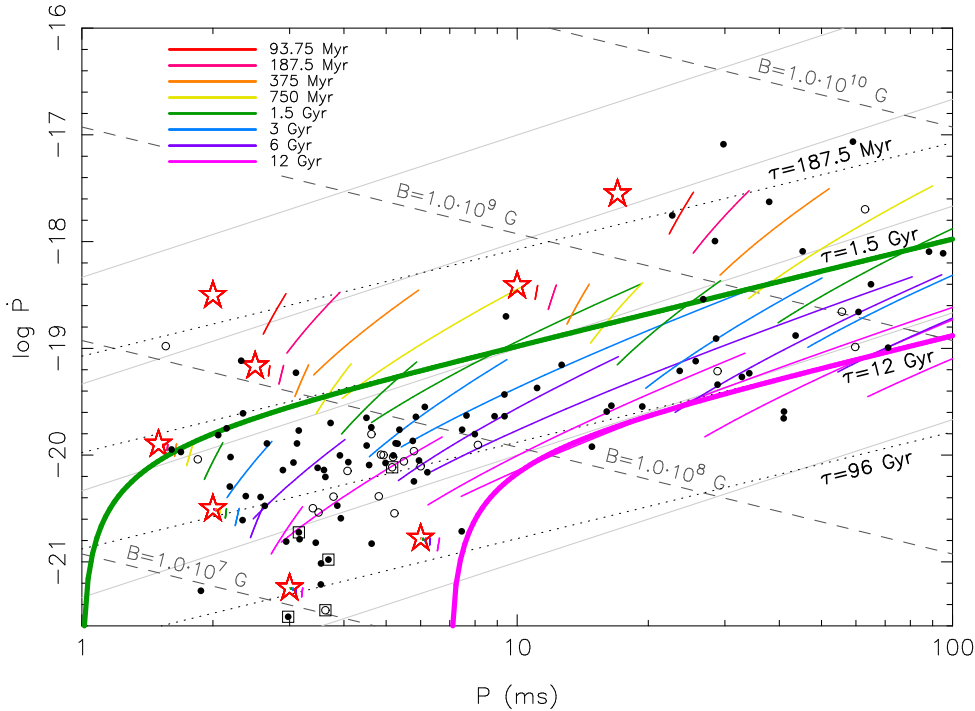} 
  \caption{Isochrones of eight hypothetical recycled pulsars born at the locations of the red stars.
             The isochrones were calculated for different values of the braking index, $2\le n\le 5$.
             Also plotted are inferred $B$-field values (dashed lines) and
             characteristic ages, $\tau$ (dotted lines).
             The thin gray lines are spin-up lines with $\dot{M}/\dot{M}_{\rm Edd}=1,\,10^{-1},\,10^{-2},\,10^{-3}$ and $10^{-4}$ (top to bottom,
             and assuming $\sin \alpha =\phi =\omega_{\rm c}=1$).
             In all calculations we assumed a pulsar mass of $1.4\,M_{\odot}$.
             It is seen how the banana shape of the two fat isochrones (see text)
             fits very well with the overall distribution of observed pulsars in the Galactic disk.
             Binary pulsars are marked with solid circles and isolated pulsars are marked with open circles, using data from
             the {\it ATNF Pulsar Catalogue} and corrected for kinematic (Shklovskii) effects.
             (Fig. adapted from Tauris et al.\ 2012.)
            } 
    \label{fig2}
 \vspace{1mm}
 \end{center}
 \end{figure}
In order to investigate if we can understand the distribution of MSPs in the $P\dot{P}$--diagram we have traced the evolution
of eight hypothetical, recycled MSPs with different birth locations ($P_0,\dot{P}_0$). 
In each case we traced the evolution as a function of age, $t$ for a constant braking index $2\le n \le 5$ and calculated 
isochrones by integration for each pulsar given that $P(t,n,P_0,\dot{P}_0)$.
The results are shown in Fig.~\ref{fig2} together with observed data.
Furthermore, we plotted two isochrones (see fat green and pink lines) calculated 
for ($P_0=1.0\,{\rm ms}$, $n=3$, $t=1.5\,{\rm Gyr}$) and
($P_0=7.0\,{\rm ms}$, $n=3$, $t=12\,{\rm Gyr}$), respectively, and 
with no restrictions on $\dot{P}_0$ (or $B_0$).

A number of interesting conclusions can be drawn from this diagram.
The overall distribution of observed pulsars
follows nicely the banana-like shape of the two fat isochrones, see 
also \cite[Kiziltan \& Thorsett~(2010)]{kt10}, and hence pulsars are recycled with a wide range of final B-fields.
Although these curves are not an attempt for a best fit to the observations
it is interesting to notice that close to 90~\% of all recycled pulsars (even up to $P=100\;{\rm ms}$) are compatible with
being born (recycled) with an initial spin period of $P_0=1-7\;{\rm ms}$ and having ages between 1.5 and 12 Gyr.
However, from a binary evolution point of view 
many of the $P=20-100\;{\rm ms}$ pulsars (the mildly recycled pulsars) are born with such relatively slow spins 
(see Section~\ref{sec:mass-spin}) and hence they need not be that old. 
This can, for example,
be verified by cooling age determinations of their WD companion stars. 
Pulsars with small values of the period derivative, $\dot{P}\simeq 10^{-21}$ hardly evolve at all in the diagram over a Hubble time. 
This trivial fact is important since it tells us that these pulsars were basically born with their currently observed
values of $P$ and $\dot{P}$ (first pointed out by \cite[Camilo~et~al.~1994]{ctk94}). Hence, some pulsars with characteristic ages
of $\tau \simeq 100\;{\rm Gyr}$ could in principle have been recycled very recently 
-- demonstrating the unreliability of $\tau$ as a true age indicator.
It is also interesting to notice PSR~J1801$-$3210 (discovered by \cite[Bates~et~al.~2011]{bates+11}) which
must have been recycled with a relatively slow birth period, $P_0 \sim \! 7\;{\rm ms}$ despite its low B-field $<10^8\;{\rm G}$,
see Fig.~\ref{fig1} for its location in the $P\dot{P}$--diagram.

\end{document}